\documentclass[conference,a4paper]{IEEEtran}

\usepackage{cite}
\usepackage{amsmath,amssymb,amsfonts}
\usepackage{algorithmic}
\usepackage{graphicx}
\usepackage{textcomp}
\usepackage{xcolor}
\usepackage{balance}
\def\BibTeX{{\rm B\kern-.05em{\sc i\kern-.025em b}\kern-.08em
    T\kern-.1667em\lower.7ex\hbox{E}\kern-.125emX}}
\begin{document}

\title{Indoor Wireless Signal Modeling with Smooth Surface Diffraction Effects}

\author{\IEEEauthorblockN{Ruichen Wang}
\IEEEauthorblockA{\textit{ECE Department} \\
\textit{University of Maryland, College Park}\\
Maryland, United States \\
rwang92@umd.edu}
\and
\IEEEauthorblockN{Samuel Audia}
\IEEEauthorblockA{\textit{CS Department} \\
\textit{University of Maryland, College Park}\\
Maryland, United States \\
sjaudia@umd.edu}
\and
\IEEEauthorblockN{Dinesh Manocha}
\IEEEauthorblockA{\textit{CS and ECE Departments} \\
\textit{University of Maryland, College Park}\\
Maryland, United States \\
dmanocha@umd.edu}
}

\author{\IEEEauthorblockN{
Ruichen Wang\IEEEauthorrefmark{1},   
Samuel Audia\IEEEauthorrefmark{2},   
Dinesh Manocha\IEEEauthorrefmark{1}\IEEEauthorrefmark{2},    
}                                     
\IEEEauthorblockA{\IEEEauthorrefmark{1}
Department of Electrical and Computer Engineering, University of Maryland, College Park (UMD), Maryland, United States}
\IEEEauthorblockA{\IEEEauthorrefmark{2}
Department of Computer Science, University of Maryland, College Park (UMD), Maryland, United States}
 \IEEEauthorblockA{Email: rwang92@umd.edu}
}

\maketitle

\begin{abstract}

 We present a novel algorithm that enhances the accuracy of electromagnetic field simulations in indoor environments by incorporating the Uniform Geometrical Theory of Diffraction (UTD) for surface diffraction. This additional diffraction phenomenology is important for the design of modern wireless systems and allows us to capture the effects of more complex scene geometries. Central to our methodology is the Dynamic Coherence-Based EM Ray Tracing Simulator (DCEM), and we augment that formulation with smooth surface UTD and present techniques to efficiently compute the ray paths. We validate our additions by comparing to analytical solutions of a sphere, method of moments solutions from FEKO, and ray-traced indoor scenes from WinProp. Our algorithm improves shadow region predicted powers by about $5dB$ compared to our previous work, and captures nuanced field effects beyond shadow boundaries. We highlight the performance on different indoor scenes and observe 60\% faster computation time over WinProp.

\end{abstract}

\vskip0.5\baselineskip

\section{Introduction}

The increased complexity of indoor wireless systems has required the development 
of more complex simulation tools to accurately design and test them.
Earlier simulation methods used simple line-of-sight
models~\cite{D. B. Green} and  current methods tend to improve the accuracy by using multipath bounces and diffraction~\cite{M. Jacob}. Typical methods for high-fidelity electromagnetic field modeling are intensive in both memory and computation time~\cite{A. F. Peterson}. Large-scale, complex scenes, such as realistic indoor spaces, use ray tracing solutions to capture fields in a reasonable amount of time and computation. 

Ray optical techniques are limited by the fact that they
cannot accurately capture edge diffraction, smooth edge diffraction, or higher
order effects such as cavity responses due to their assumption that a surface
interaction is local and planar~\cite{P. Pathak3}. Edge diffraction is typically computed using the
Uniform Geometrical Theory of Diffraction (UTD)~\cite{P. Pathak3} or Physical Theory of Diffraction
(PTD)~\cite{R. G.} which superimposes solutions for edge diffraction onto solutions of lower order effects. The latter method requires the expensive operation of integrating currents throughout surfaces in the scene and, as such, is not combined with pure geometric optics solutions. Therefore, we focus on UTD, which also allows for the accurate capture of smooth edge diffraction on Perfect Electrical Conductor (PEC) \cite{P. Pathak2} and PEC bodies coated with a thin layer of 
dielectric \cite{A. G. Aguilar} solely by tracing additional rays.

{\bf Main Results:} We present an approach that augments the Dynamic coherence-based EM Ray Tracing Simulator (DCEM)~\cite{Wang}\cite{R. Wang} that can handle complex dynamic scenes efficiently, with off surface source and receiver smooth surface diffraction modeled using UTD~\cite{P. M. Johansen,P. H. Pathak}. This theory has long been understood and used, but its usage in indoor simulation has been mostly limited to edge diffraction. Our formulation has been motivated by the importance of smooth edge diffraction for accurately capturing the real-world geometry of buildings~\cite{S. Deng}. Moreover, our approach for smooth edge diffraction can also handle human obstacles, which can be reasonably approximated by PEC cylinders~\cite{M. Ghaddar}. 
 The novel components of our work include:

\begin{enumerate}
    \item Incorporation of smooth surface diffraction effects into a general, scalable, and dynamic ray tracing simulation offering a more comprehensive tool for indoor wireless system design and testing.
    \item Rigorous validation against analytical solutions and commercial tools demonstrating the simulations marked improvements in accuracy, performance runtime, and efficiency.
    \item Our approach can enable efficient dynamic scene modeling in crowded spaces.
\end{enumerate}

We validate our additions by first comparing them to two small objects: a sphere and a metal tea cup. We compare our solution for a sphere with the analytical scattering solution, while the teacup is compared to the commercial tool FEKO's \cite{Feko} method of moments simulation. We also compare a room-scale simulation containing a PEC sphere to show the extension of our work to larger objects or scenes, and the results show clear improvements in terms of received powers ($5dB$ on average compared to the original DCEM w/o surface diffraction predictions) in the shadow region. For the third test, we compare the heat maps of the room computed by our method to the ones produced by DCEM and WinProp, where we observe clear received power raised close to shadow boundaries while maintaining the prediction accuracy outside the shadow region. We also compare the running time and efficiency and observe significant saving in running time of around $60\%$ compared to WinProp at default settings. Additionally, we demonstrate contributions beyond the shadow boundary on the smooth surface as well as attenuated fields in the lit region, as described by the UTD theory. 

\section{Uniform Geometrical Theory of Diffraction}

\subsection{Limitations of Geometric Optics}

Geometric optics is an asymptotic high-frequency approximation of the propagation of electromagnetic energy \cite{R. G.}. The method approximates the path of travel for electromagnetic fields as a ray representing the shortest path between two points. Upon intersecting a surface, the ray reflects and potentially refracts, splitting energy between the two directions. It is assumed that the ray travels through a homogeneous medium. That medium can, however, have an associated loss which is typically modeled by an exponentially decaying extinction term. Surface interactions are further assumed to be locally flat and infinite. At high frequencies and with smooth curvatures, this approximation matches well with measured data. Furthermore, the method provides run times orders of magnitude faster for large problem sizes compared to full-wave solvers, which require expensive, memory-intensive computations. 

For common frequencies of interest, like those for wireless communication, as well as realistic, nonsmooth geometry, the geometric optics assumption no longer holds. To make up for these shortcomings, the geometric optics field is superimposed with diffracted fields. This was originally proposed by Keller with the Geometrical Theory of Diffraction (GTD) \cite{J. B. Keller}, which accounted for diffraction effects using an extended ray model. The original GTD formulation had singularities in the transition between lit and unlit regions known as the shadow boundary. The Uniform Geometrical Theory of Diffraction \cite{R. G.} was then developed to account for these singularities and provide smooth, bounded calculations across the shadow boundary region. 

\subsection{Off Surface UTD Surface Diffraction}

Geometric optics can be extended by the Uniform Geometrical Theory of Diffraction to account for scattering along smooth convex surfaces even in regions that are not directly illuminated. We use the off surface formulation of the diffraction because we assume that transmitters and receivers are not located on scattering surfaces. UTD can largely be described by two equations: one for the lit region and one for the shadowed region. They are as follows \cite{R. G.}: 

\begin{equation}
\bar{E}(P) = \bar{E}^i(P)U^i + \bar{E}^{r}(P)U^r \label{eq:1}
\end{equation}

and

\begin{equation}
\bar{E}(P) = \bar{E}^d(P)(1-U). \label{eq:2}
\end{equation}

The $U^i$ and $U^r$ terms represent the Heaviside step function, signifying a direct path visibility 
for the incident and multipath reflected fields, respectively. $\bar{E}^i$ is the incident field directly from a transmitter at a receiver location $P$ while $\bar{E}^r(P)$ and $\bar{E}^d(p)$ represent the reflected and diffracted fields, respectively. They are given by

\begin{equation}
\bar{E}^{r}(P) = \bar{E}^i(Q)\cdot \bar{\bar{R}}\sqrt{\frac{\rho_1^r\rho_2^r}{(\rho_1^r + s^r)(\rho_2^r + s^r)}}e^{-jks^r} \label{eq:3}
\end{equation}

and

\begin{equation}
\bar{E}^d(P) = \bar{E}^i(Q_1)\cdot \bar{\bar{T}}(Q_1,Q_2)\sqrt{\frac{\rho^d}{s^d(\rho^d + s^d)}}e^{-jks^d}. \label{eq:4}
\end{equation}

The $\rho$ terms account for the surface curvature while $s$ represents the distance from a scattering point $Q$ on the surface and the receiver. $\bar{\bar{R}}$ is the UTD reflection coefficient accounting for both parallel and perpendicularly polarized field components \cite{P. H. Pathak}. $\bar{\bar{T}}$ is a transfer function that describes the amplitude and phase variation of the ray along geodesic paths from $Q_1$ to $Q_2$ as well as the diffraction at a surface point $Q_2$ \cite{P. M. Johansen}. The square root terms can be interpreted as the dispersion of the field along the geodesic path, and the exponential accounts for the phase difference between the scattering point and the receiver.

\section{The Dynamic Coherence-Based EM Ray Tracing Simulation}
\subsection{Prior work}
In previous works \cite{Wang}\cite{R. Wang}, DCEM was developed to handle large urban models and perform accurate computations in dynamic scenes with scalability to many moving objects. DCEM exploits a coherence-based framework and performs accurate computations in dynamic scenes. The underlying ray tracing algorithm models direct rays, reflected rays, diffracted rays, and scattered rays to perform EM propagation simulations. It uses three main characteristics: backward ray tracing, bounding volume hierarchy (BVH), and propagation path caching to accelerate the performance in dynamic environments \cite{Wang}. DCEM was further extended to large urban scenes with multiple receivers with spatial consistency and channel correlation to improve ray tracing simulation's efficiency and accuracy in dynamic scenarios and address its scalability \cite{R. Wang}. 

\subsection{Augmentation to DCEM for smooth surface diffraction}
In this work, we improve the DCEM's prediction accuracy in shadow regions by introducing smooth surface diffraction calculations as described in the previous section. By adapting how the field is calculated in the lit and unlit regions, the effects of convex surfaces are accounted for. This will commonly produce larger fields in shadow regions where there would otherwise be a sharp shadow boundary as well as interference effects in lit regions from encircling fields. We include those ray paths that satisfy the generalized Fermat's principle into the shadow region to improve the prediction accuracy. This method was explored in some other literature as \cite{R. A. Kipp,Z. Cong}, but only for small objects on small scales (a few $cms$).\\
In this work, when computing diffracted ray paths from backward ray tracing, we mark the diffraction surfaces, consider the sources that lie in the diffraction shadow region from the receiver’s perspective by UTD smooth surface diffraction, and then perform path validation back to the receiver as with reflection paths. Equation 4 is used to compute the field strength. While introducing more diffraction rays, this removes an erroneous discontinuity between the lit and shadowed regions as well as contributes to the field beyond the curved surface that would have previously been occluded. Path caching also helps accelerate the simulation process by restoring diffraction paths and reusing them for calculating other diffracted ray paths in the local area (adjacent frames). For this paper, we did not perform dynamic simulations, but we kept the potential for efficient BVH updating in dynamic scenes for future exploration~\cite{Wang}.
 
\begin{figure}[htbp]
\centerline{\includegraphics[scale=0.44]{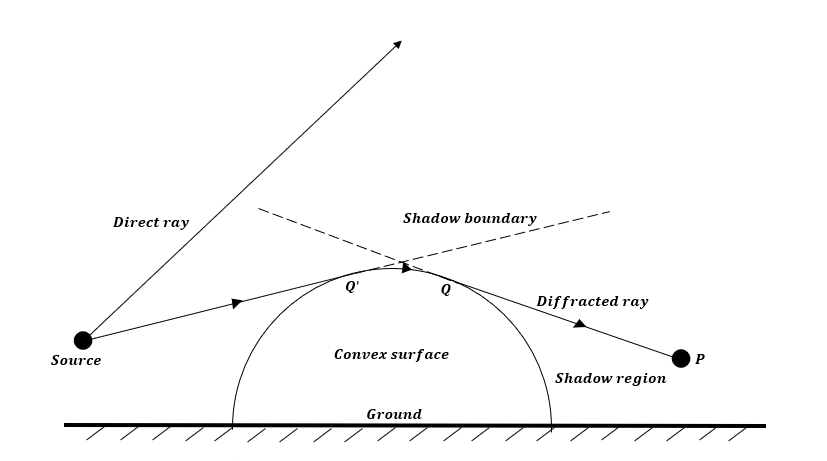}}
\caption{Diffracted ray path geometry illustration for scattering from a convex surface: some rays cast from the original source reach the convex surface and travel on the surface to further go beyond shadow boundaries. Without smooth surface diffraction, the incident rays at $Q'$ will be diffracted there. But now with smooth surface diffraction, the new paths from equivalent source $Q'$ can be generated as $Q'QP$. }
\label{fig1}
\end{figure}

In Fig.~\ref{fig1}, a typical ray path along which the electrical field propagates from the source to the field point $P$ is shown. According to GTD, the field can be expressed in terms of Equation 1 and Equation 2 in the lit and unlit regions, respectively. More detailed derivations and discussions can be found in \cite{P. Pathak}. 

\subsection{Simulation setup}
Our implementation of UTD surface diffraction was verified on small shapes such as spheres and cylinders because 
measured data for indoor spaces is unavailable and the scale makes higher fidelity simulation methods unfeasible. 
Analytical solutions were used where possible, such as for a sphere \cite{J. -M. Jin}, and Method of Moments \cite{R. F.} provided by the FEKO \cite{Feko} simulation suite was used for all other cases. Specifically, we run evaluation tests on a teacup shaped model and a sphere model, as shown in Fig.~\ref{fig10} and Fig.~\ref{fig11}.

\begin{figure}[htbp]
\centerline{\includegraphics[scale=0.37]{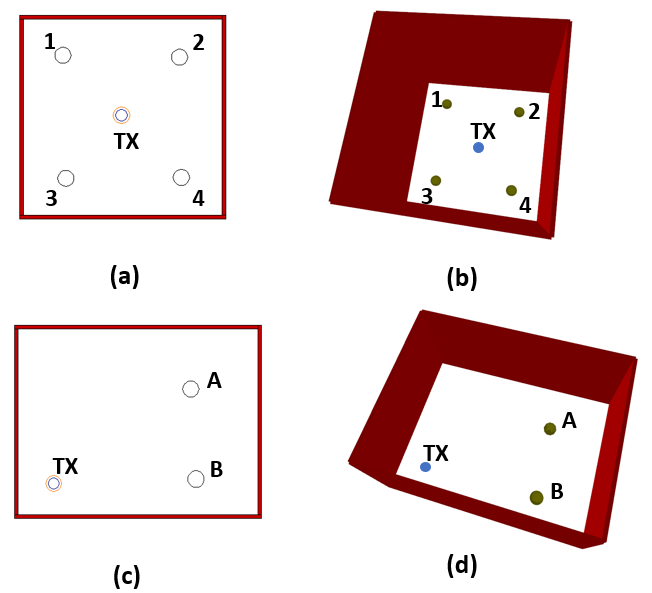}}
\caption{Comparison indoor environments. In (a), the transmitter is at the bottom left corner and the circles on the right represent metal ball positions. In (b), the positions of transmitters and spheres are displayed. In (c), the transmitter is at the center of the room and the four circles around the transmitter represent metal ball positions. In (d), the positions of transmitters and spheres are displayed. The simulated room has a floor and ceiling that was not shown in the screenshot for better illustration purposes.}
\label{fig6}
\end{figure}

We built two room-scale spaces to run comparison simulations in our ray tracing solver and the industry-standard WinProp. One comparison between DCEM w/ and w/o smooth surface diffraction calculations took place in Fig.~\ref{fig6}(a) to demonstrate the dynamic movements, and another comparison between DCEM and WinProp took place in Fig.~\ref{fig6}(c) with multiple static objects to show the performance of accuracy and efficiency. Fig.~\ref{fig6} shows the 2D and 3D views of the two simulation environments (dimensions of (a)(b): $6m*4.5m*3m$ and dimensions of (c)(d): $5m*5m*3m$). In both simulation room setups, the transmitters are placed at $0.2m$, a height similar to that of the metal balls (on the ground with a $0.2m$ radius). The transmitters are omnidirectional with the transmitting power of $40dBm$ at $2 GHz$ (wavelength $15cm$). As defined in WinProp, the circles are approximated by different numbers of line segments. In Fig.~\ref{fig6}(a)(b), the balls are approximated by 30 line segments; in Fig.~\ref{fig6}(c)(d) they are approximated by 10. The latter has fewer lines for approximation to reduce the simulation time in WinProp, with reasonable prediction outputs. The same materials are used in both simulation setups: the walls are $10 cm$ bricks and the balls have $1 mm$ metal surface.

\section{Results and discussions}

\subsection{Evaluation results from Analytical, FEKO, and DCEM solutions} 
\begin{figure}[htbp]
\centerline{\includegraphics[scale=0.26]{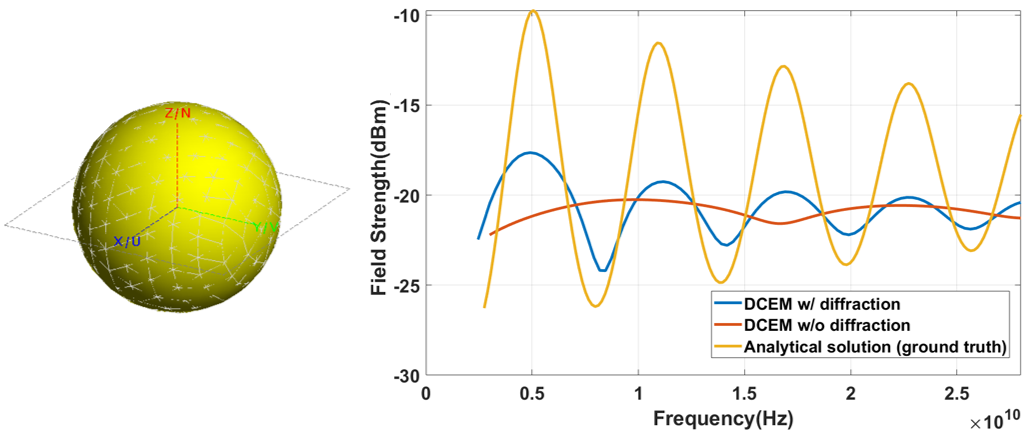}}
\caption{Left: a perfect sphere conductor with a radius of 2 cm\cite{Feko}. Right: Field strength for the sphere: the yellow line represents the analytical solution as ground truth, the red line comes from the simulation of DCEM w/o smooth surface diffraction, and the blue line from the simulation of DCEM w/ smooth surface diffraction.}
\label{fig10}
\end{figure}

\begin{figure}[htbp]
\centerline{\includegraphics[scale=0.26]{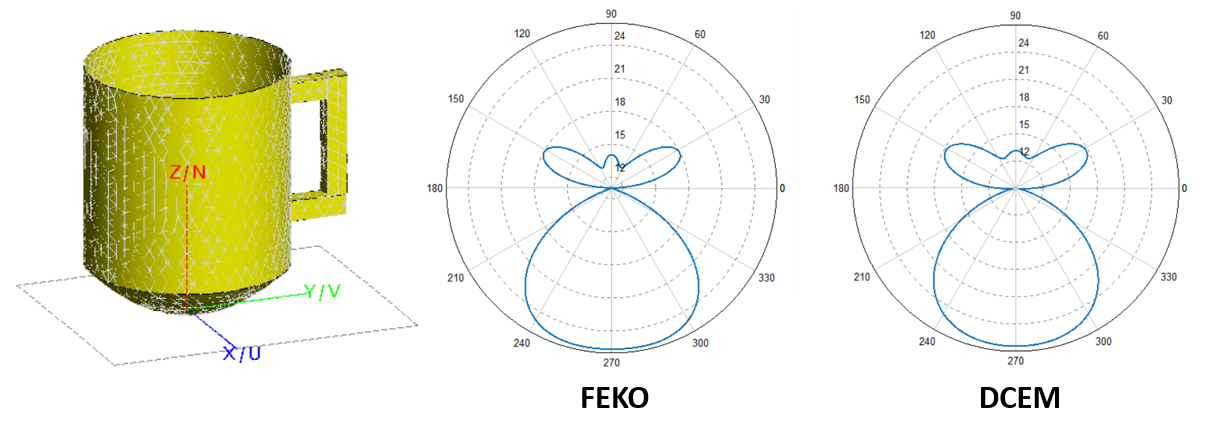}}
\caption{Left: a teacup shaped perfect conductor model with 12.5 cm height and radius of 5 cm \cite{Feko}. Middle: polar plot from FEKO. Right: polar plot from our DCEM-based method. We observe that the lower part has almost the same results as representing the teacup's cup part, while some minor differences in the upper part come from the handle part. However, it's clear that the field characteristics are perfectly identified and the minor differences might result from simulation precision settings.}
\label{fig11}
\end{figure}

In Fig.~\ref{fig10}, we show the validation field strength results of a $0.2m$ sphere computed by analytical solutions, DCEM without smooth surface diffractions, and our approach that uses DCEM with smooth surface diffractions. We see significant improvements by introducing smooth surface diffractions to reproduce the response periodicity. However, the amplitudes of peaks still show clear differences which partially result from the proximity of the smooth surface diffraction methods, but they might be aligned better in future improvements by appropriate parameter tweaks.  \\
In Fig.~\ref{fig11}, we show the teacup simulation result comparisons in polar plots. The field strength around the teacup model aligned very well and we can clearly identify the responses from the main body (lower half of the plots) and the handle part (upper half of the plots). 

\subsection{Improvements in DCEM with smooth surface diffraction}
\begin{figure}[htbp]
\centerline{\includegraphics[scale=0.5]{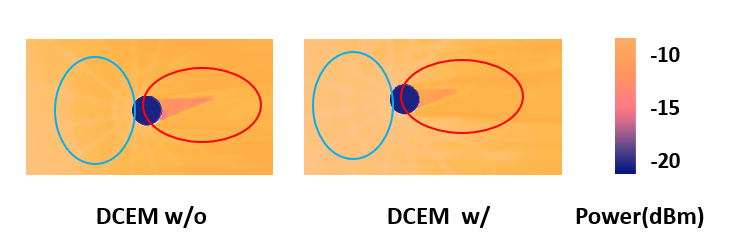}}
\caption{Heatmap comparison of DCEM without smooth surface diffraction (left) vs. DCEM with smooth surface diffraction (right) of Sphere B illustrated in Fig.~\ref{fig6}(a). We observe the same reflected rays on the front side (indicated with a blue circle); the differences mostly lie in the red circled area. This is as expected for the smooth surface diffractions that take place at the shadow boundaries and raise the field strength in the shadow region area. We explicitly show the received power along a typical path into the shadow region in Fig.~\ref{fig8}; more discussions follow.}
\label{fig7}
\end{figure}

\begin{figure}[htbp]
\centerline{\includegraphics[scale=0.3]{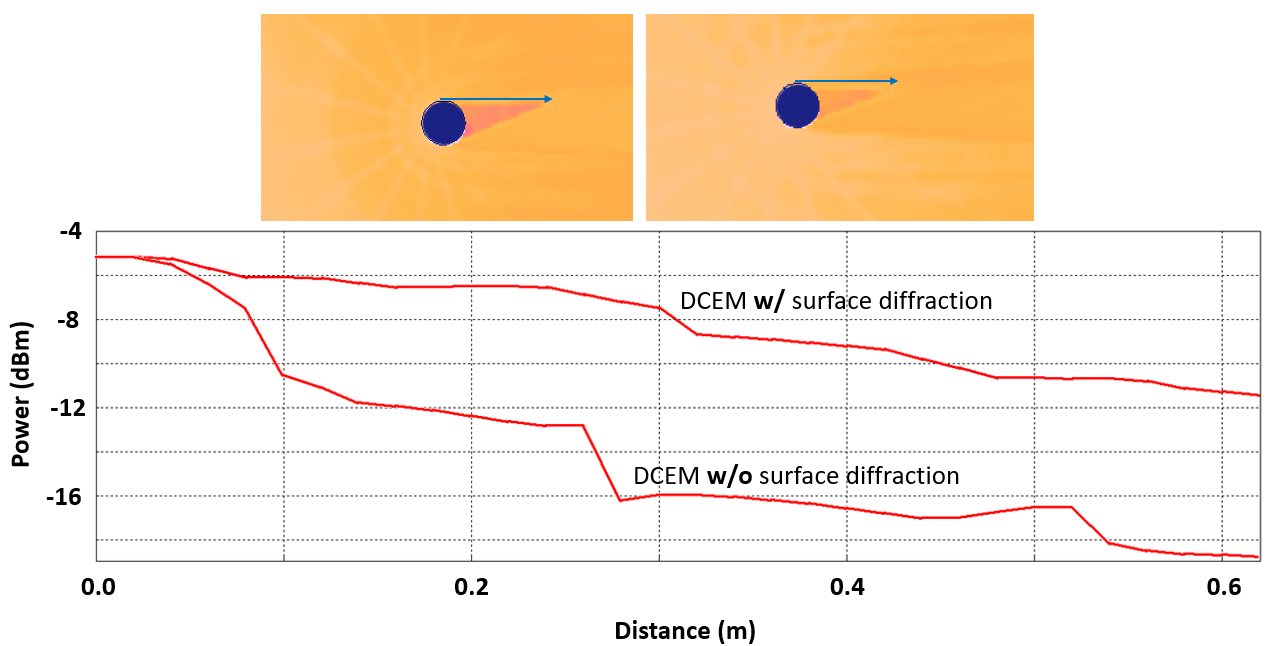}}
\caption{Received power comparison of DCEM without smooth surface diffraction vs. DCEM with smooth surface diffraction of Sphere B along a boundary path into a shadow region (indicated with the blue arrow above the received power plot). We observe the power strengths are the same at the sphere boundary and result in a $5dB$ difference at the end of the $0.62m$ path. The smooth surface diffractions contribute more significantly closer to the diffraction point on the smooth surface before $0.15m$ and decay afterward. However, as indicated in the heatmap Fig.~\ref{fig7}, the received powers should align again when the shadow region created by the metal ball blockage ends.}
\label{fig8}
\end{figure}

In this section, we show the improvements in DCEM simulations with smooth surface diffraction. The setups are discussed in Section III.C. In Fig.~\ref{fig7}, we show a snapshot of the heatmap focused on Sphere B, where smooth surface diffractions take place. While there are also diffracted rays around Sphere A, we selected Sphere B for a clearer and simpler comparison. In Fig.~\ref{fig8}, we show received powers along a typical ray path by smooth surface diffraction. With smooth surface diffraction, DCEM becomes better at shadow region field strength calculations.

\subsection{Results comparison between WinProp and DCEM}
In this section, we show the simulation results in WinProp and DCEM to show better shadow region prediction accuracy, and also compare the running times of WinProp and DCEM under such settings. The WinProp settings are specified below. We enabled the ray optical propagation model with a 3D ray tracing algorithm: the max number of transmissions, reflections, and diffractions are 3, 2, and 1, respectively. We used Fresnel Coefficients for transmission and reflections and GTD/UTD for diffractions. The TX frequency is $2GHz$, and the resolution is set to $0.02m$. The other setups are the same as discussed in Section III.C.

\begin{figure}[htbp]
\centerline{\includegraphics[scale=0.4]{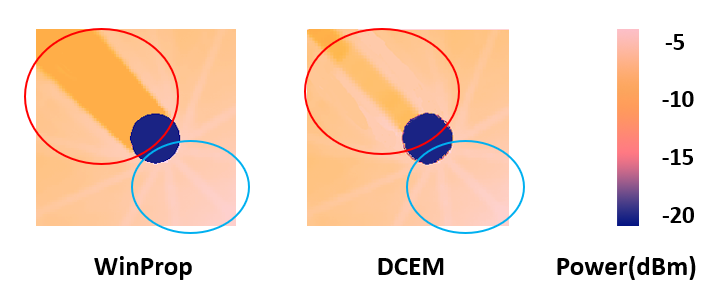}}
\caption{Heatmap comparison of WinProp (left) vs. DCEM (right) of Sphere 1, illustrated in Fig.~\ref{fig6}(c). The propagation paths are the same in the blue circled area, where most reflections happen. In the red circled area, we observe a stronger field and a thinner shadow region in the DCEM heatmap (on the right). We explicitly show the received power along a line across the shadow region to demonstrate the prediction differences in Fig.~\ref{fig9}. }
\label{fig3}
\end{figure}

\begin{figure}[htbp]
\centerline{\includegraphics[scale=0.4]{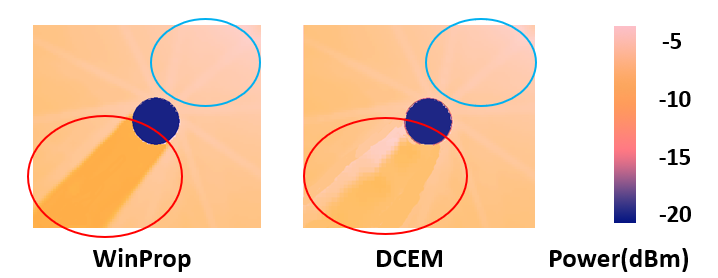}}
\caption{Heatmap comparison of WinProp (left) vs. DCEM (right) of Sphere 3, illustrated in Fig.~\ref{fig6}(c), a confirmation comparison to the observations and conclusions made in Fig.~\ref{fig3}. Again, the blue circled area shows the same reflection calculations, and the red circled area indicates stronger shadow region strength predictions in general. }
\label{fig4}
\end{figure}

\begin{figure}[htbp]
\centerline{\includegraphics[scale=0.32]{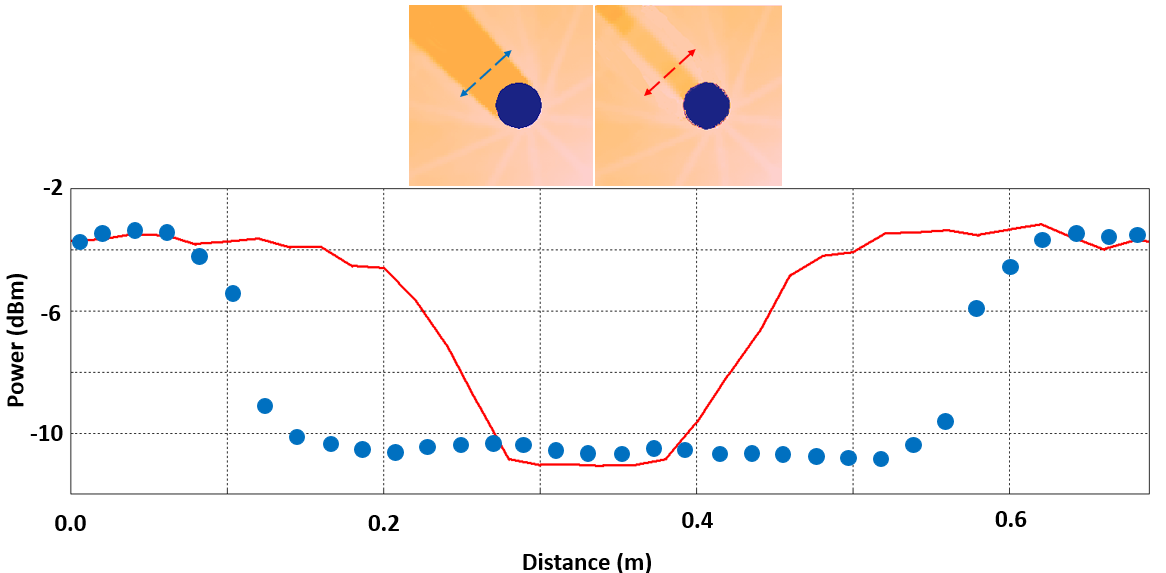}}
\caption{Received power comparison along a line across the shadow region bonds in WinProp (the left heatmap, the path indicated in a blue dashed line, results in blue dots,), and DCEM (the right heatmap, the path indicated in a red dashed line, results in a red line). We see the width of the shadow region in WinProp simulations (blue dots) is approximately $0.4m$ (bottom to bottom) while that of the DCEM results (red lines) is around $0.1m$. This is expected because the smooth surface diffraction enables more illuminated areas around the original shadow boundary regions, decaying fast after a wavelength ($0.15m$ in our settings). In addition, we see the powers at the diffraction point and outside of the region align well (shown as the powers at the two sides of this plot). This also means that the surface diffraction shall not distort other propagation path simulations.}
\label{fig9}
\end{figure}

In Fig.~\ref{fig3} and Fig.~\ref{fig4}, we show heatmap comparisons between WinProp and DCEM at Sphere positions 1 and 3, as indicated in Fig.~\ref{fig6}(c). Since WinProp does not offer smooth surface diffraction calculations, we observe clear shadow region boundaries at both Sphere positions. DCEM with smooth surface diffraction shows more power in the shadow region through diffraction around the diffraction points. We observed the same differences in the results of the other two sphere positions, 2 and 4 (not included in this paper). In Fig.~\ref{fig9}, we show the received power along a line across the shadow region in both WinProp and DCEM to demonstrate how many improvements the  smooth prediction introduced by reducing the shadow region and keeping other simulations unaffected.

In the next Table I, we show the running time recorded in three test runs under the same setups as specified earlier. We see a significant savings in running time of around $175min$ compared to WinProp at default settings. We believe that this discrepancy could be explained by redundant rays tracing repeated paths resulting in a longer running time.

\begin{table}[htbp]
\caption{Running time comparison of WinProp and our approach based on DCEM. We highlight the speedups.}
\begin{center}
\begin{tabular}{|c|c|c|c|}
\hline
\cline{2-3} 
\textbf{Simulator} & \textbf{\textit{Runtime test1}}& \textbf{\textit{Runtime test2}}& \textbf{\textit{Runtime test3}}\\
\hline
WinProp&  305min& 300min &302min\\
\hline
DCEM& 132min& 125min &124min\\
\hline
\end{tabular}
\label{tab}
\end{center}
\end{table}

\section{Conclusion and Future work}

We proposed the addition of UTD Surface Diffraction effects to indoor wireless characterization. Comparisons of our
simulation to WinProp show the inclusion of additional diffraction effects that more accurately represent the underlying physics; however, validation against measured data would still be beneficial. We would also like to explore how our approach scales to larger, more complex indoor environments and how it could handle dynamic simulations of crowds using the PEC cylinder approximation. We would also like to evaluate the accuracy by comparing with real-world measurements.

\newpage

\end{document}